\documentstyle[titlepage]{article}
\newcommand{\R}{{\rm I\kern-2pt R}}

\newtheorem{proposition}{Proposition}[section]

\newtheorem{remark}{Remark}[section]
\newtheorem{definition}{Definition}[section]

\begin{document}
\begin{center}
{\bf On the Exponentials of Some Structured Matrices\\} 
Viswanath Ramakrishna $\&$ F. Costa\\ 
Department of Mathematical Sciences
and Center for Signals, Systems and Communications\\
University of Texas at Dallas\\
P. O. Box 830688\\
Richardson, TX 75083 USA\\
email: vish@utdallas.edu 
\end{center}

\begin{abstract}
This note provides explicit techniques to compute the exponentials
of a variety of structured $4\times 4$ matrices. The procedures are fully
algorithmic and can be used to find the desired exponentials in closed
form. With one exception, they require no spectral information about the matrix
being exponentiated. They rely
on a mixture of Lie theory and one particular Clifford Algebra isomorphism.
These can be extended, in some cases, to higher dimensions when combined
with techniques such as Given rotations.

\end{abstract}

\section{Introduction}
Finding matrix exponentials is arguably one of the most
important goals of mathematical physics. In full generality,
this is a thankless task, \cite{dubious}. However, for matrices
with structure, finding exponentials ought to be more tractable. In
this note, confirmation of this phenomenon is given for a large class of
$4\times 4$ matrices with structure. These include skew-Hamiltonian,
perskewsymmetric, bisymmetric (i.e., simultaneously
symmetric and persymmetric -e.g., symmetric, Toeplitz),
symmetric and Hamiltonian etc., Some of the 
techniques presented extend almost verbatim 
to some families of complex matrices (see 
Remark (\ref{complexify}), for instance). Since such  matrices arise 
in a variety 
of physical applications in both {\it classical and quantum physics},   
it is interesting that their exponentials can be calculated
algorithmically (these lead to closed form formulae), for the most part, 
without any auxiliary information about their spectrum. For general
symmetric matrices, however, 
the spectral decomposition of a $3\times 3$ matrix
is needed [see, {\it however}, iii) of Remark (\ref{minipolyi})].
On the other hand, this spectral decomposition can itself be
produced in closed form. Thus, even for such matrices the techniques
described here can be justifiably called closed form methods.
For brevity, this note only records explicit algorithms for finding
these matrix exponentials - the resultant final formulae can easily
be written down once the reported procedures are implemented.

The methods discussed below are of two types. 
The first, which is more versatile, relies on an algebra isomorphism
of real $4\times 4$ matrices with $H\otimes H$.
This algebra isomorphism,
known from the theory of Clifford algebras and
which ought to be widely advertised, was used in a series of interesting  
articles by Mackey et al., \cite{ni,nii,niii} for
finding eigenvalues of some of the structured matrices discussed here.
The present note can be seen as a contribution of a similar type.
It is emphasized that for the preponderance of the matrices, considered
here, {\it this algebra isomorphism alone is needed}.
In particular, in this note
no use is made of any of the structure preserving rotations used in
\cite{ni,nii,niii} ever - see ii) 
of Remark (\ref{svd}).        
The second 
is based on the observation that several ``covering" space Lie group
homomorphisms, when made explicit, contain in them a recipe for finding
exponentials of matrices belonging to certain Lie algebras. This circumstance
renders the exponentiating of some $4\times 4$ matrices (real/complex)
equivalent to the job of finding the exponential of $2\times 2$ matrices
- which can be done in closed form. This method is, however, applicable
only to a limited family of matrices. Therefore, this method is presented
in an appendix.

It is worth noting that, though most of the
structured matrices considered here were chosen for their importance in
applications, 
the real enabling structure is that present in $H\otimes H$. This is
especially illustrated by certain normal matrices
[see Definition (\ref{special})].

The balance of this note is organized as follows. In the next section
some notation and one observation which is used throughout is recorded.
In the same section the relation between $H\otimes H$ and $gl(4, R)$ is 
presented. 
The third section 
discusses a wide family of matrices which can be exponentiated using
the aformentioned algebra isomorphism. The final section offers conclusions.
In the first appendix the second approach to exponentiating matrices in
$p(4, R)$ and $so(2,2, R)$ (see next section for notation) 
is presented in a manner that makes the connection to the covering
space homomorphism between $SU(2)\times SU(2)$ and $SO(4, R)$ explicit
(see Remark (\ref{unified})).
In the final section thirteen classes of matrices are listed
which can be exponentiated
by mimicking verbatim two situations studied earlier.

In closing this introductory section it is  
noted that by combining these techniques with techniques such as structure
preserving similarites, e.g, Givens rotations, \cite{recipe}, 
one can extend these results, in many cases, to find
algorithmically the exponentials of structured matrices of size {\it bigger
than four}. In other words, one can use such similarities (normally used
in the literature for reduction to canonical forms) to reduce the exponential
calculation to dimension four or lower of matrices with similar structure.
In principle, this would provide closed form formulae for
the exponentials of such structured matrices, since one can explicitly
write down the desired Givens type similarities.
However, it is more accurate to say that this
implies an algorithmic procedure for exponentiating such matrices.
For matrices for which this is possible (e.g., symmetric matrices), 
the details of this procedure
is routine and hence will not be pursued here.

\section{Notation and Preliminary Observations}
The following definitions and notations will 
be frequently met in this work: 
\begin{itemize}
\item $gl(n, R)$ and $gl(n,C)$ represent the real (resp. complex)
$n\times n$ matrices.
\item $sl(n, R)$ and $sl(n, C)$ represent the real (resp. complex)
traceless matrices. $SL(n,R)$ and $SL(n, C)$ represent the real (resp. complex)
matrices of determinant one.
\item $SU(n)$ represents  the $n\times n$ unitary matrices of determinant
one. $su(n)$ represents the $n\times n$ skew-Hermitian, traceless matrices.
\item $R_{n}$ represents the matrix with $1$ on the anti-diagonal and
zeroes elsewhere. $p(n,R)$ and $p(n, C)$ represent the $n\times n$ real
(resp. complex) matrices, $A$, satisfying $A^{T}R_{n} + R_{n}A = 0$.
These matrices are also said to be {\it perskewsymmetric}.     
{\it Persymmetric} matrices are those matrices, $X$, which satisfy
$X^{T}R_{n} = R_{n}X$. $P(n, R)$ (respectively $P(n, C)$) is the
set of matrices (real/complex), $X$, which satisfy $X^{T}R_{n}X = R_{n}$.
  
\item $J_{2n}$ is the $2n\times 2n$ matrix which, in block form,  
is given by $J_{2n} = \left ( \begin{array}{cc}
0_{n} & I_{n}\\
-I_{n} & 0_{n}
\end{array}
\right )$. $sp(2n, R)$ and $sp(2n, C)$ represent those real (resp. complex)
$2n\times 2n$ matrices which satisfy $X^{T}J_{2n} + J_{2n}X = 0$. 
Such matrices are also called {\it Hamiltonian}. Matrices, $Z$, satsifying
$Z^{T}J_{2n} = J_{2n}Z$ are called {\it skew-Hamiltonian}.
\item $I_{p,q} = \left ( \begin{array}{cc}
I_{p} & 0\\
0 & -I_{q}
\end{array}
\right )$. $so (p, q, R)$ and $so(p, q, C)$ represent the real (resp. complex)
$n\times n$ matrices ($n = p + q$), $X$, satisfying
$X^{T}I_{p,q} + I_{p,q}X = 0$.
\item The {\it anti-trace} of an $n\times n$ matrix is the sum of the elements
on its anti-diagonal. $X$, $n\times n$, is {\it  anti-scalar}
if $X = \gamma R_{n}$,
with $\gamma \in R$ (or $C$).
  
\item Throughout $H$ will be denote the {\it quaternions}, 
while $P$ stands for the
{\it purely imaginary} quaternions, tacitly identified with $R^{3}$.   

\end{itemize}

\begin{remark}
\label{minipolyi}
\begin{itemize}
\item i) {\rm Throughout this note, use of the following observation will
be made: Let $X$ be an $n\times n$ matrix satsifying $X^{2} + c^{2}I_{n}
= 0, c\neq 0$. Then $e^{X} = \cos (c) I_{n} + \frac{\sin (c)}{c}X$.
Here $c^{2}$ is allowed to be complex, and $c$ is then taken to be
$\sqrt{r} e^{i\frac{\theta}{2}}$, with $c^{2} = re^{i\theta},
\theta \in [0, 2\pi)$.}
 
\item {\rm Occassionally the fact that any matrix which satisfies $X^{3} = 
-c^{2}X, c\neq 0$,
satisfies $e^{X} = I + \frac{\sin (c)}{c}X + \frac{1-\cos (c)}{c}X^{2}$
(the Rodrigues's formula) will also be used.}

\item iii) {\rm Explicit formulae 
for $e^{A}$ can be produced if the minimal polynomial
of $A$ is known and it is low in degree 
(cf., \cite{zenii} where such formulae are written down from
the characteristic polynomial). Since it is possible
to find the minimal polynomial of many of the matrices considered
here explicitly (i.e., without any spectral information),
this removes the need for
the spectral decomposition, mentioned in the introduction, for $A$ symmetric.
However, since the corresponding
explicit formulae for $e^{A}$ are more complicated than the
ones in i) and ii), they will not be pursued here. See the
conclusions for an illustration of this issue.}

\end{itemize}
\end{remark}    

\vspace*{2.7mm}
                                                                                
\noindent {\bf $H\otimes H$ and $gl(4, R)$}:
$\bullet$ Associate to each product
tensor $p\otimes q\in H\otimes H$,
the matrix, $M_{p,q}$, of the map which sends $x\in H$
to $px\bar{q}$, identifying $R^{4}$ with $H$ via this basis $\{1,i,j,k\}$,
Extending this to the full tensor product by linearity,
yields an algebra isomorphism
This connection, which
is known from the theory of Clifford Algebras, has been put
to great practical use in solving eigenvalue problems for structured
matrices by Mackey et al., \cite{ni,nii,niii}.
 It can also be used for finding exponentials,
$e^{A}, A\in gl(4, R)$ via the following procedure:

\vspace*{2.7mm}
                                                                                
\noindent {\bf General Algorithm for $e^{A}$ Using $H\otimes H$}
                                                                                
\begin{itemize}
\item I) Identify $u\in H\otimes H$, corresponding to $A$ via
this isomorphism.
\item II) Find $e^{u}\in H\otimes H$ (in general, this will be possible
in closed form only if $u$ (and, hence $A$)
possesses additional structure).
\item  III) Find the
matrix $M$ corresponding to $e^{u}$ 
- this is $e^{A}$.
\end{itemize}

\noindent {\it Note:} Throughout this work, tacit use of $H\otimes H$
representations of matrices in $gl(4, R)$ will be made. 
These can be easily obtained from the
entries of the $4\times 4$ matrix in question (see 
\cite{ni,nii,niii} for some instances). 
In particular, $R_{4} = M_{j\otimes i}, J_{4} = M_{1\otimes j}$.

\section {Exponentials of Structured $4\times 4$ Matrices}
In this section the algebra isomorphism between $H\otimes H$
and real $4\times 4$ matrices
will be used to find exponentials of various structured matrices.
For many of these matrices, their exponentials can be found directly
from their $H\otimes H$ representations. These will be presented first.
For the remaining the singular value factorization of matrices,
no bigger than $3\times 3$, is needed. This can be done in closed
form, \cite{symmetricsing}. These will be presented next.

\subsection {Exponentials Directly From $H\otimes H$ Representation}

Below a (by no means exhaustive) list of nine families of
real $4\times 4$ matrices, whose exponentials can be directly found from
their $H\otimes H$ representations, is presented.  
These families seem to be ubiquitous
in applications.

\begin{enumerate}
\item {\bf $4\times 4$ skew-symmetric matrices:}
The corresponding element
in $H\otimes H$ is $p\otimes 1 + 1\otimes q$ with $p, q \in P$. 
For finding its exponential, it is noted that
$p\otimes 1$ and $1\otimes q$ commute, so the exponential of the
sum is the product of the individual exponentials.
Now consider Consider $(p\otimes 1)^{2}  =
 -(\mid\mid p \mid\mid^{2})1\otimes 1$. Thus, $p\otimes 1$ is annihilated
by a quadratic polynomial, and its exponential is therefore is
$[\cos (\mid\mid p \mid\mid )1 + \frac{\sin (\mid\mid p \mid\mid )}
{\mid\mid p \mid\mid )}p]\otimes 1 = x\otimes 1$, with $x$ a unit
quaternion. Likewise, $e^{(1\otimes q)}
= 1\otimes y$, $y$ a unit quaternion. Thus, $e^{A}$ is the matrix
$M_{x\otimes y}$ - which is a different point of view on the 
$SU(2)\times SU(2), SO(4, R)$ relation. 

\item {\bf $4\times 4$ perskewsymmetric matrices:}
Such matrices $P$ have $H\otimes H$ representations 
$p\otimes i + \alpha (j\otimes 1) + j\otimes q
+ \beta (1\otimes i) = X + Y + Z + W$ with $p\in {\mbox span} \ \{i,k \},
q \in {\mbox span} \ \{j,k\}, \alpha , \beta \in R$,
we find $X,Y$  both commute with
each of $Z$ and $W$. Hence $e^{P} = e^{(X+Y)}e^{(Z+W)}$. 
Further, $XY=-YX, ZW=-WZ$.

\noindent Next, since $(X+Y)^{2} = 
(\mid\mid p\mid\mid^{2} + \alpha^{2}) 1\otimes 1$,
$e^{(X+Y)} = \cosh (\lambda_{1})(1\otimes 1) +\frac{\sinh (\lambda_{1})}
 {\lambda_{1}}(X+Y)
= \cosh (\lambda_{1})(1\otimes 1) +\frac{\sinh (\lambda_{1})}
{\lambda_{1}} (p\otimes i + \alpha (j\otimes 1))$, with $\lambda_{1}
=\sqrt{(\mid\mid p\mid\mid^{2} + \alpha^{2})}$.
Likewise, $e^{(Z+W)} =
\cosh (\lambda_{2})(1\otimes 1) +\frac{\sin (\lambda_{2})}
 {\lambda_{2}}(Z+ W)
= \cosh (\lambda_{2})(1\otimes 1) +\frac{\sinh (\lambda_{2})}
{\lambda_{2}} (j\otimes q + \beta (1\otimes i))$, with
$\lambda_{2} = \sqrt{(\mid\mid q\mid\mid^{2} + \beta^{2})}$.

Hence, $e^{P}$ is the matrix representation of
$\{ \cosh (\lambda_{1})(1\otimes 1) +\frac{\sinh (\lambda_{1})}
{\lambda_{1}} [(p\otimes i) + \alpha (j\otimes 1)]\} \{
\cosh (\lambda_{2})(1\otimes 1) +\frac{\sinh (\lambda_{2})}
{\lambda_{2}} [(j\otimes q + \beta (1\otimes i)]\}$.
 
\item {\bf $4\times 4$ skew-Hamiltonian Matrices:}
Such matrices, $S$, have $H\otimes H$ representations of the form
$b (1\otimes 1) + p\otimes j + 1\otimes (ci + dk)$, with
$b,c,d\in R$ and $p\in P$. Clearly the $b (1\otimes 1)$ component
commutes with the remaining summands. Thus $e^{S} = e^{b}
{\mbox exp} (p\otimes j + 1\otimes (ci + dk))$.  
Now, $(p\otimes j + 1\otimes (ci + dk))^{2} =
- (-\mid\mid p \mid\mid^{2} + c^{2} + d^{2})(1\otimes 1)$.
Indeed the two summands 
summands anti-commute, while $(p\otimes j)^{2} = \mid\mid p \mid\mid^{2}
(1\otimes 1); (ci + dk))^{2} = - (c^{2} + d^{2})(1\otimes 1)$. 
Hence $e^{S} = e^{b} (\cos (\lambda ) (1\otimes 1) +
\frac{\sin (\lambda )}{\lambda} (p\otimes j + 1\otimes (ci + dk)))$,
with $\lambda = \sqrt{ (-\mid\mid p \mid\mid^{2} + c^{2} + d^{2})}$.
Note $\lambda \in C$.

\item {\bf Five Jordan Algebras:} See Appendix II 

\item {\bf Eight Lie Algebras:}
See Appendix II. {\it In particular, one member of this list is precisely
$so(2,2,R)$.} 

\item {\bf Simultaneously Hamiltonian, Symmetric, Persymmetric Matrices:}
These have       
$H\otimes H$ representations of the form 
$M = X + Y + Z = \beta (j\otimes i) + \gamma (i\otimes k) +
\delta (k\otimes k), \beta , \gamma , \delta\in R$. 
Now $X$ commutes with both $Y, Z$ while
$Y, Z$ anti-commute, and each of $X, Y, Z$ squares to a positive 
constant times $1\otimes 1$. Hence $e^{M}$ is the matrix representation of 
\[
 e^{X}e^{(Y+Z)} = [\cosh (\beta )(1\otimes 1)
+ \sinh (\beta ) (j\otimes i )]
[\cosh (\lambda )(1\otimes 1) 
+ \frac{\sinh (\lambda )}{\lambda} (\gamma (i\otimes k) +
\delta (k\otimes k)], \lambda = \sqrt {\gamma^{2}
+ \delta^{2}}\]

\item {\bf Some Symmetric Toeplitz Matrices: }
The general case of a symmetric, Toeplitz matrix is subsumed by
the case of bisymmetric matrices - see Remark (\ref{bisymmetric}) below.
Here we identify two important classes which 
do not require the intervening spectral
factorization calculations for the general case. 
\begin{itemize}
\item {\it Symmetric, Toeplitz, Tridiagonal Matrix}:
Since such a matrix is met frequently in applications, it worth noting
that its exponential can be directly computed in closed form.
Indeed, their $H\otimes H$ representations are given 
by $a(1\otimes 1) + \frac{b}{2} (j\otimes i) + \frac{b}{2}(i\otimes j)
+ b(k\otimes j), a,b \in R$. Expressing this as $X + Y+ Z+ W$,
we see $X$ and $Y$ commute with both $Z, W$ and further $XY=YX, ZW = -WZ$.
Hence $e^{(X + Y+ Z +W)} = e^{X}e^{Y}e^{(Z+W)} = 
e^{a}[\cosh (\frac{b}{2})(1\otimes 1) +\sinh (\frac{b}{2}) 
(j\otimes i)][\cosh (c) (1\otimes 1) + \frac{\sinh (c)}{c}
( (i + k)\otimes j)], c = \frac{\sqrt{5}}{4}b$.      

\item {\it Symmetric, Toeplitz Matrix $S$ Satisfying $s_{13} = 0$}:
This implies that the {\it second superdiagonal and subdiagonal vanish}.
Such matrices have $H\otimes H$ representations of the form $a(1\otimes 1) +
b(j\otimes i) +  c(i\otimes j) + b(k\otimes j)$. Now, the first and second
summand commute amongst themselves and with the remaining summands. While
the third and the fourth anti-commute.
Hence,
\[
e^{S} =e^{a} [\cosh (b)(1\otimes 1) + \sinh (b)(j\otimes i)]
[\cosh (\lambda )(1\otimes 1) + \frac{\sinh (\lambda )}{\lambda}
(c(i\otimes j) + b (k\otimes j))], \lambda = \sqrt{b^{2} + c^{2}}
\]
 
\end{itemize}

\item {\bf Certain Normal Matrices:} The general case of normal
matrices is subsumed by the algorithm below for a symmetric matrix,
since the case of skew-symmetric matrices has already been dealt with
(a matrix is normal iff its symmetric and skew-symmetric parts commute).
Here we discuss a subclass 
which does not require the spectral factorization
calculations needed for exponentiating a symmetric matrix. 
This subclass is described via the following:
\begin{definition}
\label{special}
{\rm Consider a normal $N = S + T$, with $S$ its symmetric part and
$T$ its skew-symmetric part. Expressing $T$ as the sum of two commuting
skew-symmetric matrices, $T_{1} = M_{s\otimes 1}$ and $T_{2}
= M_{1\otimes t}, s,t\in P$ 
it is assumed that $\mid\mid s\mid\mid \neq \mid\mid t\mid\mid$.
Such matrices will be called} special normal.
\end{definition}
Note that special normality forces $T\neq 0$. Special normality also
implies that $[S, T_{i}] = 0, i=1,2$ (this will be shown below).
It is this condition that makes exponentiation in closed form possible. 

Indeed, consider, first the case that $T_{1}\neq 0$.
Letting $ S = a (1\otimes 1)
+ p\otimes i + q\otimes j + r\otimes k$, the assumption $[S, T_{1}]
= 0$ forces, in conjunction with the linear independence of the
elements $e_{x}\otimes e_{y}, e_{x}, e_{y} = i, j, k$, {\it each} of the 
$p\otimes i, q\otimes j, r\otimes k$ to commute with $s\otimes 1$.
This implies that each of $p, q, r$ is parallel to $s$ and hence
the symmetric part of $N$ can be expressed succinctly as
\[
a(1\otimes 1) + s\otimes \hat{t}, s, \hat{t}\in P
\] Now the condition,
$[S, T_{2}] = 0$ forces $t$ to be parallel to $\hat{t}$.
Hence we find
\[
 e^{N} = [e^{a}(\cosh (\lambda )I_{2} + \frac{\sinh (\lambda )}{\lambda} 
(s\otimes \hat{t})] (e^{s}\otimes 1)(1\otimes e^{t}),
\lambda = \mid\mid s\mid\mid\mid\mid \hat{t}\mid\mid
\] 
If $T_{1} = 0$, then the condition $[S, T_{2}] = 0$ implies that each of
$p,q,r$ are parallel to one another (w.l.o.g $p\neq 0$), and hence
$S = p\otimes \hat{t}$,
with $\hat{t} = kt, k\in R$. Hence the above formula holds with minor
modification. 

Next, it will be shown that special normality implies $[S, T_{i}] = 0,
i = 1,2$. One first shows 
\begin{equation}
\label{quarticmini}
T^{4} + 2(\mid\mid s\mid\mid^{2} + \mid\mid t\mid\mid^{2})T^{2}
+ [(\mid\mid s\mid\mid^{2} - \mid\mid t\mid\mid^{2})^{2}]I
= 0
\end{equation}
The calculation leading to the above simultaneously shows i) $T$'s minimal
polynomial is quadratic iff either of $s$ or $t$ vanishes
(in this case, trivially $[S, T_{i}] = 0, i=1,2$; ii) $T$'s minimal polynomial
is cubic iff $\mid\mid s\mid\mid = \mid\mid t\mid\mid$. Hence,
w.l.o.g $T$'s minimal polynomial is quartic, i.e., $T$
is {\it non-derogatory}.

Next, since $S, T$ commute,
they are simultaneously diagonalizable, via some unitary
matrix $U$. Consider $U^{*}TU
= U^{*}T_{1}U + U^{*}T_{2}U$. 
The last two matrices commute (since $T_{1},
T_{2}$ commute) and their sum is diagonal. If the entries of the diagonal
matrix $U^{*}TU$ are all distinct and non-zero, then
the matrices $U^{*}T_{1}U$ and $U^{*}T_{2}U$ are themselves diagonal.
Thus they also commute with $U^{*}SU$,
which implies that $S$ commutes with
both the $T_{i}$. 
Note $T$ being non-derogatory implies the 
assumptions about the diagonal entries of $U^{*}TU$, in view of the nature of
the eigenvalues of a $4\times 4$ skew-symmetric matrix.

\item {\bf Certain Non-Toeplitz Bisymmetric Matrices} 
Every persymmetric matrix is of the form $RS$, with $S$ symmetric.
Similarly, Hamiltonian matrices 
are of the form $JS$, with $S$ symmetric.
{\it Such matrices can often be exponentiated in closed form, if in addition,
$R_{4}S = SR_{4}$ (resp. $J_{4}S = SJ_{4}$).}

\vspace*{2.7mm}
 
\noindent Indeed, since $RS = SR$, and
$R^{2} = I$, we find $e^{RS} = \cosh (S) + R\sinh (S)$ (this equation
is valid in any dimension). Now, $S = a(1\otimes 1)
+ p\otimes i + q\otimes j + r\otimes k$ satisfies $R_{4}S = SR_{4}$ iff 
i) $p$ is parallel to $j$ and ii) $q,r$ are perpendicular to $j$. 
If, in addition
we suppose either $q$, $r$ are parallel or $q,r$ are perpendicular
to one another, then exponentiation in closed form is possible.
For brevity the former possibility is assumed.
Hence
\[
S =  a(1\otimes 1) +\epsilon (j\otimes i) + (\alpha i + \beta k)\otimes
(\gamma j + \delta k)
\]  
Note, in particular, that $RS$ is symmetric, persymmetric, {\it but
not Toeplitz}.

\noindent Writing $RS$ as $R (\mu I_{4} + \tilde{S})$, with
$\tilde{S} = X + Y$, we see that it suffices to find $e^{R\tilde{S}}$.
Now, notice that $X$ and $Y$ commute and $X^{2} = \epsilon^{2}I,
Y^{2} = (\alpha^{2} + \beta^{2})(\gamma^{2} + \delta^{2})I =
\lambda^{2}I$. Hence, $\cosh (\tilde{S}) = \cosh (X)\cosh (Y) + 
\sinh (X)\sinh (Y)$, and $\sinh (\tilde{S}) 
= \sinh (X)\cosh (Y) + \sinh(Y)\cosh(X)$.
But $\sinh (X) = \frac{\sinh (\epsilon )}{\epsilon}X;
\sinh (Y) = \frac{\sinh (\lambda )}{\lambda}Y; \cosh (X) =\cosh (\epsilon )I;
\cosh (Y) = \cosh (\lambda )I$. Hence $e^{RS}$ is the matrix given by:
\[
[\cosh (\mu )I_{4} + \frac{\sinh (\mu)}{\mu}R]
[\cosh (\epsilon)\cosh (\lambda) I + \frac{\sinh (\epsilon )\sinh (\lambda )}
{\lambda\epsilon}XY] + R[\frac{\sinh (\epsilon )\cosh (\lambda )}{\epsilon}X 
+ \frac{\sinh (\lambda )\cosh (\epsilon )}{\lambda}Y] 
\]

\vspace*{2.7mm}
\noindent Similarly, if $JS = SJ$, one finds (since $J^{2} = -I$) that
\[
e^{JS} = \cos (S) + J_{2n}\sin (S)
\]
Now if $S$, symmetric, commutes with
$J$, then fortunately (or unfortunately) $J_{4}S$ is also simultaneously  
skew-symmetric, and therefore the previous formula is yet another way
of exponentiating $J_{4}S$. Hence, the details are omitted.

\end{enumerate}

\subsection {The General Symmetric Case}
Exponentiating the general $4\times 4$ symmetric matrix requires the 
spectral factorization of a $3\times 3$ matrix (which can be done in closed
form). Before getting to that, the principal enabling feature
of the algorithm below
is described by the following:

\begin{proposition}
\label{anticommute}
{\rm The exponential of $a(1\otimes 1) + \sum_{i=1}^{3}u_{i}\otimes v_{i},
u_{i}, v_{i}\in P$, with $\{u_{i}, i=1,\ldots , 3\}, 
\{v_{i}, i=1, \ldots , 3\}$ each an orthogonal triple in $R^{3}$ is
given by $e^{a}\Pi_{i=1}^{3}e^{(u_{i}\otimes v_{i})}$, with
$e^{(u_{i}\otimes v_{i})} = \cosh (\mid\mid u_{i}\mid\mid
\mid\mid v_{i}\mid\mid) (1\otimes 1) +
 \frac{\sinh (\mid\mid u_{i}\mid\mid
\mid\mid v_{i}\mid\mid)}{(\mid\mid u_{i}\mid\mid 
\mid\mid v_{i}\mid\mid)}(u_{i}\otimes v_{i})$.}  
\end{proposition} 

{\it Proof:} It suffices to observes that each of the summands
in $a(1\otimes 1) + \sum_{i=1}^{3}u_{i}\otimes v_{i}$ commutes with
each other due to the orthogonality property. The formula for
$e^{(u_{i}\otimes v_{i})}$ is now just a consequence of 
$(u_{i}\otimes v_{i})$ squaring to a positive constant times
the identity.

\begin{remark}
{\rm 
If the triples 
$ \{u_{i}\}, \{v_{i}\}$ were instead each parallel to each other,
then once again the exponential of  
$a(1\otimes 1) + \sum_{i=1}^{3}u_{i}\otimes v_{i}$ is quickly computed,
since now once again each summand commutes with one another. 
There are other possible configurations which will render the calculation
of the exponential in closed form too. However, these will not be pursued
here for brevity.}
\end{remark}

\begin{remark}
\label{svd} 
\begin{itemize}
\item i) {\rm Consider the element $p\otimes i + q\otimes j + r\otimes k,
p,q,r\in P$. Then, as observed in \cite{ni}, if 
$\sum_{i=1}^{3}\sigma_{i}u_{i}v_{i}^{T}, u_{i}, v_{i}\in R^{3}$ is
the singular value factorization
of the real $3\times 3$ matrix, $[p \mid q \mid r]$,
it follows that  $p\otimes i + q\otimes j + r\otimes k
=
\sum_{i=1}^{3}\sigma_{i}u_{i}\otimes v_{i}$,
where the vectors $u_{i}, v_{i}$ have been
identified with the corresponding pure quaternions
(in lieu of the elegant proof in \cite{ni},
one can also verify this via direct calculations which show that if for 
$p_{i},q_{i},r_{i},s_{i}\in P, i=1,\ldots , 3$,
the $3\times 3$ matrices $\sum_{i=1}^{3}p_{i}q_{i}^{T}, 
\sum_{i=1}^{3}r_{i}s_{i}^{T}$ coincide, then 
$\sum_{i=1}^{3}M_{p_{i}\otimes q_{i}} = \sum_{i=1}^{3}M_{r_{i}\otimes s_{i}}$). 
Since the
$\{u_{i}\}, \{v_{i}\}$ are each an orthonormal triple, the exponential
of $p\otimes i + q\otimes j + r\otimes k$, which equals the
exponential of $\sum_{i=1}^{3}\sigma_{i}u_{i}\otimes v_{i}$, can be
explicitly found
by using Proposition (\ref{anticommute}). The only issue is computing
the singular value factorization of a real $3\times 3$ matrix. However,
this is the spectral factorization of a real $3\times 3$ symmetric
matrix, which itself can be done in closed form, \cite{symmetricsing}.
It is interesting to note that the technique described
in \cite{symmetricsing}, consisting of $3\times 3$ matrix manipulations,
can itself be implemented via quaternions. } 

\item ii) {\rm Note the subsequent
rotations employed in \cite{ni} to diagonalize
a symmetric matrix are} not required, {\rm since diagonalization
is not being employed here to compute exponentials. Only the reduction
to form used in Proposition (\ref{anticommute}) is needed.} 

\end{itemize}
\end{remark}

\noindent This leads to the following algorithm for the
{\it exponential of a $4\times 4$ symmetric matrix}:

\begin{itemize}
\item Represent the matrix as $a(1\otimes 1) +  
p\otimes i + q\otimes j + r\otimes k,
p,q,r\in P$.
\item Compute the singular value factorization,
$\sum_{i=1}^{3}\sigma_{i}u_{i}v_{i}^{T}, u_{i}, v_{i}\in R^{3}$ of
the real $3\times 3$ matrix, $[p \mid q \mid r]$.
\item Compute the exponential of $ a(1\otimes 1) + \sum_{i=1}^{3}
\sigma_{i}u_{i}\otimes v_{i}$ via Proposition (\ref{anticommute}).
The $4\times 4$ matrix representing this element of $H\otimes H$ is
$e^{A}$.
\end{itemize} 
\begin{remark}
\label{bisymmetric}
{\rm The special classes of $4\times 4$ bisymmetric
matrices (i.e., simultaneously
symmetric and persymmetric) and 
$4\times 4$ symmetric and Hamiltonian matrices
are, of course, subsumed by the foregoing algorithm. However, it is
worth pointing out, in view of their importance in applications, that the
singular value factorization needed is easier to find than in the
fully symmetric case. Indeed, a bisymmetric matrix is
represented by $a(1\otimes 1) + b(j\otimes i) + p\otimes j + q\otimes k,
p, q\in {\mbox span} \ \{i,k\}, a, b\in R$. Thus, it suffices to find
the singular value factorization of the $2\times 2$ matrix $[p\mid q]$
- which is the spectral factorization of a $2\times 2$ real symmetric
matrix. Likewise, a symmetric, Hamiltonian matrix is represented by
$q\otimes i + r\otimes k$. Thus, it suffices to find the singular value
factorization of the $3\times 2$ matrix $[ p \mid q]$ (only two of
the left singular vectors are needed). There are many other cases of
symmetric matrices possessing additional symmetry which are susceptible
to the same observation.}
\end{remark}   

\begin{remark}
\label{complexify}
Extension to Complex Matrices: {\rm Some of the procedures extend
to special classes of complex matrices. This is illustrated for
matrices in $so(4, C)$. Such a matrix can be represented in
the form $\alpha_{1}M_{i\otimes 1} + \beta_{1}M_{j\otimes 1}
+ \gamma_{1}M_{k\otimes 1} + \alpha_{2}M_{1\otimes i}
+ \beta_{2}M_{1\otimes j} + \gamma_{2}M_{1\otimes k}
= \sum_{l=1}X_{l}$, with $\alpha_{i},\beta_{i},\gamma_{i}\in C$.
Now the fact that these constants are complex does not prevent
each of $X_{1},\ldots , X_{3}$ from commuting with each of
$X_{4}, \ldots , X_{6}$. Neither does it prevent each of
$X_{1},\ldots , X_{3}$ anti-commuting with one another nor
each of $X_{4}, \ldots , X_{6}$ anti-commuting with one another.
Finally, $X_{i}^{2} = -c_{i}^{2}I_{4}$ for each $i=1, \ldots , 6$,
for some $c_{i}\in C$. Hence the exponential is given
by \[
[\cos (\lambda_{1})I_{4} + \frac{\sin (\lambda_{1})}{\lambda_{1}}
(\alpha_{1}M_{i\otimes 1} + \beta_{1}M_{j\otimes 1}
+ \gamma_{1}M_{k\otimes 1}][\cos (\lambda_{2})I_{4}
+ \frac{\sin (\lambda_{2})}{\lambda_{2}}
(\alpha_{2}M_{1\otimes i}
+ \beta_{2}M_{1\otimes j} + \gamma_{2}M_{1\otimes k})]
\]
with $\lambda_{i}^{2} = - (\alpha_{i}^{2} + \beta_{i}^{2} 
+ \gamma_{i}^{2}), i=1,2$.  
Similarly the technique for $p(4, R)$ extends verbatim
to $p(4, C)$. However, while the methods based on the singular value
factorization extend verbatim for purely imaginary symmetric matrices,
they are not applicable to general complex symmetric matrices.
To see what is needed for the extension, consider traceless symmetric
matrices (w.l.o.g). Let $A_{R}$ and $A_{I}$ be the real and imaginary
parts of $A$. Since these are symmetric as well, one can associate
two triples $(p_{i},q_{i},r_{i})\in P^{3}, i=1,2$. Let $M_{i}
= [p_{i}\mid q_{i}\mid r_{i}]$ be the corresponding real $3\times 3$
matrices. If these could be simultaneously brought into the
canonical forms $M_{i} = \sum_{k=1}^{3}\sigma_{k}^{i}u_{k}v_{k}^{T}$,
with the $u_{k}$ and $v_{k}$ orthonormal, $\sigma_{k}^{i}\in R$,
then clearly the algorithm for real symmetric matrices would extend
verbatim to such matrices. Many sufficient conditions are known for
such simultaneous canonical form, \cite{gib}. One such condition
is that both $M_{1}M_{2}^{T}, M_{2}^{T}M_{1}$ should be symmetric.} 
\end{remark}

\section{Conclusions}
In this note, closed form formulae are provided for exponentials
of several important families of real (and complex) $4\times 4$ matrices.
In conjunction, with techniques such as Givens rotations, these formulae
provide algorithms for exponentiating classes of structured matrices in
higher dimensions. The principal technique is the invocation of the
associative algebra isomorphism between $gl (4, R)$ and $H\otimes H$. 
It is the ease of multiplication in $H\otimes H$ which facilitates the
discovery of closed form exponentials for many matrices    

It is possible
to write down exponentials of matrices once their minimal polynomial
is known (especially if they are at most quartic). 
However, these formulae
themselves can be quite complicated and hence they were not pursued in 
this note. This is exemplified by generic $4\times 4$ skew-symmetric matrices,
whose minimal polynomial is quartic. The corresponding exponential formula,
though
equivalent to the one given here, is substantially more complicated. 
In our opinion most $4\times 4$ matrix calculations
should be done in $H\otimes H$. The formulae for the minimal polynomial
of a $4\times 4$ skew-symmetric matrix
[see Equation (\ref{quarticmini})], without any spectral information,
is yet another vivid illustration.    

Clearly, $\tilde{H}\otimes \tilde{H}$ is associative algebra isomorphic
to $gl(4, c)$, where $\tilde{H}$ is the complexification of $H$.
One can identify the latter with $gl(2,C)$. However, it is better to
view its elements as $ q =
x_{0} + x_{1}i + x_{2}j + x_{3}k, x_{i}\in C$ and define
$\bar{q} = \bar{x_{0}} - \bar{x_{1}}i - \bar{x_{2}}j - \bar{x_{3}}k$.
This notion of conjugation is equivalent to Hermitian conjugation
in $gl(4, C)$. This does not, however, render calculating exponentials
in $su(4)$ as simple as in $so(4, R)$ (after all one cannot run away
from the curse of dimensionality by such an artifice). However, several
Hermitian and skew-Hermitian matrices (e.g., 
whose real and imaginary parts come
from special normal real matrices) are easily exponentiated.  

\section{Appendix I:}
In this appendix, a different approach to the exponentiation
of matrices in $p(4, R), so(2,2, R), p(3, R)$ is described,
which reduces the problem
to the exponentiation of $2\times 2$ matrices (this is equally applicable
to their complex counterparts). This is first illustrated for matrices
in $so(3, R)$ and $so(4, R)$ since this should be reasonably well known
terrain. Attention, in particular, is drawn to Remark (\ref{unified}),
which provides the correct heurisitics needed to generalize 
this to the matrices
in $p(4, R), so(2,2, R), p(3, R)$.

Consider an element $A\in so(3, R)$.
Its exponential can be computed explicitly via the Rodrigues
formula. The usual derivation of this relies on the fact that $A$ satisfies
\[
A^{3} + \lambda^{2} A = 0, \lambda\in R
\]
                                                                                
Any matrix which satisfies this equation will satisfy the Rodrigues
formula. There is a equally well-known relation between $su(2)$ and
$so(3, R)$. What is, perhaps, less appreciated is that this relation
yields an explicit technique to find $e^{A}, A\in so(3, R)$. 
To describe this, fix a $G\in SU(2)$. Consider
$V = \{ A \mid A^{*} = A, {\mbox Tr}(A) = 0\}$. $SU(2)$ acts
via conjugation on elements $A\in V$, viz., $\phi_{G}(A) = GAG^{-1}$. It is
well known, that upon identifying $V$ with $R^{3}$ through
the basis $\{ \sigma_{k} , k=x,y,z\}$, this action yields a proper
rotation of $R^{3}$. Thus, we get a homomorphism,  $\phi : SU(2)
\rightarrow SO(3, R)$, which sends $G$ to the matrix of $\phi_{G}$ with
respect to the basis $\{ \sigma_{k} , k=x,y,z\}$.
This is a surjective, two-one, homomorphism. Linearizing this map, we
get a Lie-algebra isomorphism $\psi: su(2)\rightarrow so(3, R)$, viz.,
$\psi (A)$ is the matrix of the linear map which sends $v\in V$
to $Av-vA$ {\it with respect to
the} $\{ \sigma_{k} , k=x,y,z\}$ {\it basis}.
with $A\in su(2)$. This is a Lie-algebra isomorphism. From elementary
considerations in Lie theory $\psi$ and $\phi$ provide the following
technique to find $e^{A}, A\in so(3, R)$:
                                                                                
\vspace*{2.7mm}
\begin{itemize}
\item i) Find $B = \psi^{-1}(A)$ in $su(2)$
\item ii) Compute $e^{B}\in SU(2)$
- this can be explicitly done since satisfies the condition
in i) of Remark (\ref{minipolyi})., 
\item iii) Compute the matrix $\phi_{e^{B}}$ - this is $e^{A}$.
\end{itemize}
                                                                                
This is arguably easier to use than the Rodrigues formula (it is left to
the reader to verify that the two result in the same formula). This is
not to disparage the Rodrigues formula - it applies to situations where Lie
theory would have no visible role. But the fact that a $3\times 3$
exponential has been computed with a $2\times 2$
calculation is significant. Similar and even better savings occur
by such arguments.
                                                                                
\noindent {\bf Exponentials in $so(4, R)$:} 
There is a well known two-one Lie group
homomorphism denoted by
$\phi: SU(2)\times SU(2) \rightarrow SO(4, R)$, given by the action
of $SU(2)\times SU(2)$ on the vector space, $V$, of real linear combinations
of $I_{2}, i\sigma_{k}, k= x,y,z$, viz., for fixed $G,H\in SU(2)\times SU(2)$,
let $\phi_{G,H} V\rightarrow V$ be given by $\phi_{G, H}(X) =
GXH^{-1}, X\in V$. Once again this is a proper rotation of
$R^{4}$ (identified with $V$ via this basis), and $\phi (G, H)$ is precisely
the {\it matrix of this map with respect to this basis}. Linearizing
this gives a Lie algebra isomorphism, $\psi : su(2)\times su(2)
\rightarrow so(4, R)$ which sends $(X,Y)\in su(2)\times su(2)$ to
the matrix of the map (with respect to the $I_{2},i\sigma_{k}$ basis)
which sends $Z\in V$ to $XZ-YZ$. 
This yields an algorithm to find $e^{A}, A\in so(4, R)$, {\it which reduces
to finding two $2\times 2$ exponentials in $su(2)$} - the statement
of the algorithm is omitted (mimick the $p(4, R)$ algorithm given below).
                                                                                
The corresponding relations between $SL(2,C)$
(respectively $SL(2,C)\times SL(2,C)$) and $SO(3,C)$ (respectively $SO(4, C)$)
once again reduce exponentiation of matrices in $so(3, C)$ and $so(4, C)$
to $2\times 2$ calculations.
Note that the fact that
$SO(3, C)$ etc., are not compact does not matter for the veracity of
this procedure. All that is needed for finding $e^{A}$
is that the corresponding $\phi$
be a Lie group homomorphism (it need not even be surjective)
and the corresponding $\psi$ be a Lie algebra
isomorphism.

\begin{remark}
\label{unified}
{\rm Traditional proofs of the $SU(2)$ covering
of $SO(3, R)$ proceed by i) using $su(2)$ itself as the vector space
$V$, and ii)then, by constructing a bilinear form, $K(X, Y)
= {\mbox Tr} ({\mbox ad}\ X
\ {\mbox ad}\ Y)$ on $su(2)$ and showing that this is preserved by
the action of $SU(2)$. For our purposes it is more useful to
proceed differently. On any (sub)space of
$2\times 2$ matrices, there are two obvious candidates for quadratic
forms, viz., i) ${\mbox Tr} (X^{2})$; and ii) ${\mbox det} (X)$.
One is even lead inexorably to these forms upon inspecting
the forms of the maps $\phi$
used above for both $so(3, R)$ and $so(4, R)$. Polarizing these two
leads to the following choices:
\begin{equation}
\label{trace}
L_{1}(X, Y) = {\mbox Tr} (XY)
\end{equation}
                                                                                
\begin{equation}
\label{determinant}
L_{2}(X, Y) = {\mbox det} (X + Y) - {\mbox det} (X) - {\mbox det} (Y)
\end{equation}

It is easy to see that, with the choice of bases made in the derivation
of the $so(3, R)$ (resp. $so(4, R)$) algorithms, that the symmetric
matrices representing these two forms are precisely $2I_{3}$
(resp. $I_{4}$). This immediately shows that the matrix of the
corresponding $\phi$'s are orthogonal.}  
\end{remark}

\begin{remark}
Lorenz Lie Algebra: {\rm Here a different perspective on the
work of \cite{zeni} on the exponentials of matrices in $so(1,3, R)$
is provided. Indeed, letting $V$
be the $R$-linear span of $\{I_{2}, \sigma_{x}, \sigma_{y}, \sigma_{z}\}$
(i.e., $V$ is the space of $2\times 2$ Hermitian matrices), it 
is found that the matrix
of $L_{2}(X,Y)$ is precisely $2I_{1,3}$. If $SL(2,C)$ acts on $V$
via $\phi_{M}(v) = MvM^{*}, v\in V, M\in SL(2,C)$, then $L_{2}(X,Y)$
is preserved and the matrix of $\phi_{M}$ in this basis is in the Lorenz group.
Linearizing $\phi$, we get a technique to find exponentials in $so(1,3, R)$,
cf., \cite{zeni}}.
\end{remark}

Below the same thinking is used to compute exponentials in $p(4, R)$,
$so(2,2, R)$ and $p(3, R)$. The method can be applied to several other
Lie algebras stemming from symmetric, non-degenerate, bilinear forms
on $R^{4}$. However, we limit ourselves to these cases for brevity.

\vspace*{2.7mm}

\noindent {\bf Exponentials in $p(4, R)$:}

Consider $gl(2, R)$, identified with $R^{4}$
via the basis, $\{ E_{11}, E_{12}, -E_{21}, E_{22}\}$. Let $SL(2, R)\times
SL(2, R)$ act on $gl(2, R)$, via $\phi_{G, H}(X) = GXH^{-1}$.
This action leaves the bilinear form $L_{2}(X, Y)$ of
Equation (\ref{determinant}) invariant. Furthermore the symmetric matrix
representing it, with respect to this basis, is precisely $R_{4}$.
Thus the matrix of $\phi_{G, H}$ is in $P(4, R)$.
Linearizing this we get a Lie-algebra isomorphism (that this is a Lie-
algebra homomorphism is standard - it is
easily verified that it is
an isomorphism): $\psi: sl(2, R)\times sl(2, R)\rightarrow p(4, R)$,
which sends a pair $(g,h)\in sl(2, R)\times sl(2, R)$ to the matrix
of the linear map $L_{g,h}(X) = gX-Xh, X\in gl(2, R)$
with respect to the $\{ E_{11}, E_{12}, -E_{21}, E_{22}\}$ basis.
This leads to the following algorithm to find $e^{A}, A\in p(4, R)$.
                                                                                
\vspace*{2mm}
\noindent {\it Algorithm for $e^{A}, A\in p(4, R)$:}
\begin{itemize}
\item i) Find the pair $(g,h) =
\psi^{-1}(A)\in
sl(2, R)\times sl(2, R)$;
\item ii) Find $G = e^{g}, H = e^{h}$.
This is easily done since $g, h$ satisfy the equation
in Remark (\ref{minipolyi} -i))
\item iii) Find the matrix of $\phi_{G, H}$ with respect to the above basis.
This is $e^{A}$.
\end{itemize}

\vspace*{2.7mm}

\noindent {\bf Exponentials in $so(2,2, R)$:}

Now
identify $gl(2, R)$ with $R^{4}$ via the basis $\{ I_{2}, E_{12} - E_{21},
\sigma_{x}, \sigma_{z}\}$. Then the matrix of $L_{2}(X, Y)$, with respect to
this basis is precisely $I_{2,2}$. Let $SL(2, R)\times SL(2,R)$ act on
$gl(2, R)$ via $\phi_{G, H}(V) = GVH^{-1}, G, H \in SL(2, R),
V\in gl(2, R)$. Then $\phi_{G, H}$ preserves $L_{2}(X, Y)$ and hence its
matrix, with respect to this basis, is in $SO(2,2, R)$. Linearizing this
action we get a Lie algebra isomorphism $\psi: sl(2, R)\times
sl (2, R) \rightarrow so(2,2, R)$, with $\psi (g, h)$ being the matrix
of the linear map $\psi_{g,h}(v) = gv -vh, v\in gl(2,R)$ with respect to
the same basis. This leads to an algorithm, similar to the previous one,
for finding $e^{A}, A\in so(2,2, R)$.

\vspace*{2.7mm}

\noindent {\bf Exponentials in $p(3,R)$:}

Now identify $R^{3}$ with the real span
of $E_{12}, \sigma_{x}, E_{21}$. This is $sl(2, R)$.
Then the matrix of $L_{1}(X, Y)$, 
with respect to this basis, is, upto a constant, $R_{3}$.
Let $SL(2, R)$ act on this copy of $R^{3}$ via $\phi_{G} (h) = GhG^{-1}$.
This action preserves $L_{1}(X, Y)$. Thus, the matrix of $\phi_{G}$
is in $P(3,R)$ and the map $\phi: SL(2, R)\rightarrow P(3, R)$ is easily
seen to be a Lie group homomorphism. Linearizing $\phi$ leads to
a Lie algebra isomorphism $\psi: sl(2, R)\rightarrow p(3, R)$
which sends $h\in sl(2, R)$ to the matrix of the linear map, which sends
$X\in sl(2, R)$ to $hX-Xh$ (identifying $sl(2, R)$ with $R^{3}$
via the above basis). This leads to an algorithm for finding $e^{A},
A\in p(3, R)$.

\begin{remark}
{\rm i) The last calculation can be mimicked to find exponentials in
$so(2,1, R)$. Indeed, identify $sl(2, R)$ with $R^{3}$ via the basis
$\{ \sigma_{x}, \sigma_{z}, E_{12} - E_{21} \}$ and proceed 
verbatim as in the $p(3, R)$ case. ii) All of the above calculations
extend to find exponentials in $p(4, C)$ etc., The only difference is
one works with complexifications of the various Lie algebras introduced
before, i.e., $sl(2, C)\times sl(2,C)$ for $p(4, C)$ etc.,}
\end{remark} 
 
\section{Appendix II}

In this appendix are listed i) five classes of matrices, each 
a Jordan algebra, which can be exponentiated by mimicking the
technique for skew-Hamiltonian matrices; ii) eight classes of 
matrices, each forming a Lie algebra, which can be 
exponentiated by mimicking the
technique for perskewsymmetric matrices. In most cases the 
technique extends to their complex analogues (e.g., $so(2,2,C)$), 
cf., Remark (\ref{complexify}).
In both lists, both the $H\otimes H$ representation and the $2\times 2$
block representations are provided.
\begin{remark}
{\rm Let $M_{1}, M_{2}$ be two invertible, symmetric (resp. skew-symmetric)
matrices, with the corresponding bilinear form on $R^{n}$ denoted
by $<,>_{M_{1}}, <,>_{M_{2}}$. The two forms are defined
to be equivalent if there is an
orthogonal matrix $G$ such that $G^{T}M_{1}G = M_{2}$. 
If this is the case then
the corresponding Jordan algebras, $J_{i} = \{ X\mid X^{T}M_{i} = M_{i}X\},
i=1,2$ and the corresponding Lie algebras $L_{i} =
\{ X\mid X^{T}M_{i} = -M_{i}X\},i=1,2$ are conjugate. Specifically
$J_{2} = G^{T}J_{1}G, L_{2} = G^{T}L_{1}G$. Thus, if one knows exponentials
of matrices in $J_{1}$ (resp. $L_{1}$), then one can find exponentials
of matrices in $J_{2}$ (resp. $L_{2}$) 
provided $G$ is explicitly described. 

In the first list, the first two Jordan algebras pertain to bilinear 
forms which are equivalent to $J_{4}$, 
while all the matrices in the second list
stem from symmetric forms equivalent to $R_{4}$. While it is possible to
explicitly construct the corresponding $G$'s}, it is far easier to work
with the matrices in these lists directly.
\end{remark}

\noindent {\bf Exponentials of Five Jordan Algebras}

\begin{itemize}
\item $p\otimes k + a (1\otimes 1) + 1\otimes (bi + cj), a,b, c\in R,
p\in P$. The block representation is 
$\left ( \begin{array}{cc}
A & B\\
C & D
\end{array} \right )$, with $B, C$ some $2\times 2$ scalar matrices,
$D = \left ( \begin{array}{cc}
\alpha & \beta \\
\gamma & \delta
\end{array} \right )$ and $A = \left ( \begin{array}{cc}
\delta & -\beta \\
-\gamma & \alpha
\end{array} \right )$.

\item $p\otimes i + a (1\otimes 1) + 1\otimes (bj + ck),
a,b, c\in R,
p\in P$. The $2\times 2$ block representation is 
$\left ( \begin{array}{cc}
A & B\\
C & D
\end{array} \right )$, with $A, D$ some $2\times 2$ scalar matrices,
$B = \left ( \begin{array}{cc}
\alpha & \beta \\
\gamma & \delta
\end{array} \right )$ and $C = \left ( \begin{array}{cc}
-\delta & \beta \\
\gamma & -\alpha
\end{array} \right )$.

\item $i\otimes q + a (1\otimes 1) + (bj + ck)\otimes 1,
a,b, c\in R,
q\in P$. The block representation is 
$\left ( \begin{array}{cc}
A & B\\
C & D
\end{array} \right )$, with
$A, D$ some $2\times 2$ scalar matrices,
$B = \left ( \begin{array}{cc}
\alpha & \beta \\
\gamma & \delta
\end{array} \right )$ and $C = \left ( \begin{array}{cc}
-\alpha & \gamma\\
\beta & -\delta
\end{array} \right )$.
\item $j\otimes q + a (1\otimes 1) + (bi + ck)\otimes 1,
a,b, c\in R,
q\in P$. The block representation is
$\left ( \begin{array}{cc}
A & B\\
C & D
\end{array} \right )$, with
$B, C$ some $2\times 2$ anti- scalar matrices,
$C = \left ( \begin{array}{cc}
\alpha & \beta \\
\gamma & \delta
\end{array} \right )$ and $A = \left ( \begin{array}{cc}
\alpha & -\gamma \\
-\beta & \delta
\end{array} \right )$
\item $k\otimes q + a (1\otimes 1) + (bi + cj)\otimes 1,
a,b, c\in R,
q\in P$. The block representation is
$\left ( \begin{array}{cc}
A & B\\
C & D
\end{array} \right )$, with
$B, C$ some $2\times 2$ zero-trace diagonal matrices,
$D = \left ( \begin{array}{cc}
\alpha & \beta \\
\gamma & \delta
\end{array} \right )$ and $A = \left ( \begin{array}{cc}
\delta & \beta\\
\gamma & \alpha
\end{array} \right )$.
\end{itemize}

\noindent {\bf Exponentials of Eight Lie Algebras} 
\begin{itemize}
\item $so(2,2, R)$. The $H\otimes H$ representation is
$a(1\otimes i) +p\otimes i + b(i\otimes i) + i\otimes q,
p, q\in {\mbox span} \ \{j,k\}, a, b \in R$.
The block representation is 
$\left ( \begin{array}{cc}
A & B\\
B^{T} & C
\end{array}\right )$, where $B$ is any $2\times 2$ matrix, while
$A, C$ are $2\times 2$ anti-diagonal matrices with zero anti-trace.
\item $p\otimes j + a(j\otimes 1) + j\otimes q + b(1\otimes j),
p.q\in {\mbox span} \ \{i,k\}, a, b\in R$. The
block representation
is $\left ( \begin{array}{cc}
A & B\\
B^{T} & C
\end{array}\right )$, where $B$ is any $2\times 2$ matrix, while
$A, C$ are $2\times 2$ anti-scalar matrices.

\item $p\otimes k + a(k\otimes 1) + k\otimes q + b(1\otimes k),
p,q\in {\mbox span} \ \{i, j\}, a, b\in R$. The block
representation is 
$\left ( \begin{array}{cc}
A & B\\
C & D
\end{array}\right )$, where $A, D$ are  $2\times 2$ anti-scalar matrices matrix, while
$B =
\left ( \begin{array}{cc}
\alpha & \beta \\
\gamma & \delta
\end{array} \right )$ and $C = \left ( \begin{array}{cc}
\alpha & -\gamma\\
-\beta & \delta
\end{array}\right )$.

\item $p\otimes i + a(k\otimes 1) + k\otimes q + b(1\otimes i),
p\in {\mbox span} \ \{i,k\}, q\in {\mbox span} \ \{j,k\}, a, b \in R$.
The block representation is 
$\left ( \begin{array}{cc}
A & B\\
C & D
\end{array}\right )$, with $B, C$ $2\times 2$ anti-scalar matrices, while
$A = \left ( \begin{array}{cc}
\alpha & \beta \\
\gamma & \delta
\end{array} \right )$ and $D = \left ( \begin{array}{cc}
-\alpha & \gamma \\
\beta & -\delta
\end{array}\right )$.

\item $p\otimes j + a(k\otimes 1) + k\otimes q + b(1\otimes j),
p\in {\mbox span}\ \{i,j\}, q\in {\mbox span}\ \{i,k\}, a, b\in R$.
The block representation is 
$\left ( \begin{array}{cc}
A & B\\
C & D
\end{array}\right )$,
with $A, D$ zero-trace, diagonal $2\times 2$ matrices, while
$B = \left ( \begin{array}{cc}
\alpha & \beta \\
\gamma & \delta
\end{array} \right )$ and $C = -\left ( \begin{array}{cc}
\delta & \beta \\
\gamma & \alpha
\end{array} \right )$.

\item $p\otimes j + b(i\otimes 1) + a(1\otimes j) + i\otimes q,
p\in {\mbox span}\ \{j,k\}, q\in {\mbox span}\ \{i,k\}, a, b\in R$.
The block representation is
$\left ( \begin{array}{cc}
A & B\\
C & D
\end{array}\right )$,
with $B, C$ $2\times 2$ scalar matrices, while $D =
= \left ( \begin{array}{cc}
\alpha & \beta \\
\gamma & \delta
\end{array} \right )$ and $A = \left ( \begin{array}{cc}
\delta & -\beta \\
-\gamma & \alpha
\end{array} \right )$.

\item $p\otimes k + a(i\otimes 1) + b(1\otimes k) + i\otimes q,
p\in {\mbox span}\ \{j,k\}, q\in {\mbox span}\ \{i,j\}, a, b\in R$.
The $2\times 2$ block representation is
$\left ( \begin{array}{cc}
A & B\\
C & D
\end{array}\right )$, with $B, C$ $2\times 2$ zero anti-trace, anti-diagonal
matrices, while $D = \left ( \begin{array}{cc}
\alpha & \beta \\
\gamma & \delta
\end{array} \right )$ and $A = - \left ( \begin{array}{cc}
\delta & \beta \\
\gamma & \alpha
\end{array} \right )$.

\item $p\otimes k + a(j\otimes 1) + b(1\otimes k) +
j\otimes q,
p\in {\mbox span}\ \{i,k\}, q\in {\mbox span}\ \{i,j\}, a, b\in R$.
The block representation is 
$\left ( \begin{array}{cc}
A & B\\
C & D
\end{array}\right )$, with $A, D$ $2\times 2$ diagonal, zero-trace matrices,
while $B = \left ( \begin{array}{cc}
\alpha & \beta \\
\gamma & \delta
\end{array} \right )$ and $C = \left ( \begin{array}{cc}
\delta & \beta \\
\gamma & \alpha
\end{array} \right )$.
\end{itemize}


\begin{thebibliography}{99}
\bibitem{dubious} C. Moler and C. Van Loan, {\it Siam Review}, {\bf 45},
3, (2003).
\bibitem{ni} N. Mackey, {\it Siam J. Matrix  Analysis},
{\bf 16}, 421, (1995).
\bibitem{nii} H. Fassbender, D. Mackey $\&$ N. Mackey,
{\it Linear Algebra $\&$ its Applications}, {\bf 332}, 37, (2001).
\bibitem{niii} D. Mackey, N. Mackey $\&$ S. Dunleavy,
{\it Structure Preserving
Algorithms for Perplectic Eigenproblems}, Numerical Analysis Report 427,
Manchester Center for Computational Mathematics, (2003).

\bibitem{recipe} W. H. Press, S. A. Teukolsky,
W. T. Vettering and B. P. Flannery, {\it Numerical Recipes in C},
II edition, Cambridge University Press, 1992.
 
\bibitem{zenii} A. Barut, J. Zeni $\&$ A. Laufer, {\it J. Phys A}, {\bf 27},
5239, (1994).
\bibitem{zeni} J. Zeni $\&$ W. Rodrigues, {\it Hadronic J}, {\bf 13},
317, (1990). 
\bibitem{symmetricsing} A. Bojanczyk $\&$ A. Lutoborski, {\it Siam
J. Matrix Analysis}, {\bf 12}, 41, (1991).
\bibitem{gib} P. M. Gibson, {\it Linear Algebra $\&$ its Applications},
{\bf 9}, 45, (1974).
\end{thebibliography}
\end{document}